\documentclass[12pt]{article}
\usepackage[cp1251]{inputenc}

\usepackage{amsfonts,amssymb,amsthm,mathtools}
\usepackage{amsmath}
\usepackage{icomma}
\usepackage{latexsym}
\usepackage{graphicx}
\usepackage{subfigure}
\usepackage {bm}

\usepackage{geometry} 
\begin{document}
	
	\begin{center}
		\textbf{QUANTUM PROBING OF SINGULARITIES AT EVENT HORIZONS OF BLACK HOLES}
	\end{center}

	\begin{center}
		
		{V.~P.~Neznamov\footnote{vpneznamov@mail.ru, vpneznamov@vniief.ru}}\\
		
		\hfil
		{\it \mbox{	Russian Federal Nuclear Center--All-Russian Research Institute of Experimental Physics},  Mira pr., 37, Sarov, 607188, Russia \\
			National Research Nuclear University MEPhI, Moscow, 115409, Russia} \\
	\end{center}
	
	
	

\begin{abstract}
	\noindent
	\footnotesize{It is proved that coordinate transformations of the Schwarzschild metric to 
		new static and stationary metrics do not eliminate the mode of a particle 
		''fall'' to the event horizon of a black hole. This mode is unacceptable for 
		the quantum mechanics of stationary states.}\\
	
	
	
\end{abstract}


\section{Introduction}	

In Refs. \cite{bib1}, \cite{bib2}, we 
considered the interaction of scalar particles $\left( {S=0} \right)$, 
photons $\left( {S=1} \right)$, fermions $\left( {S=1/2} \right)$ with 
Schwarzschild, Reissner--Nordstr\"{o}m, Kerr and Kerr--Newman black holes with 
zero and non-zero cosmologic constants. For the above metrics and particles 
with different spins, existence of the mode of a particle ''fall'' to event 
horizons has been found. The mode of a quantum--mechanical ''fall'' of 
particles to the singular center is, for instance , examined in detail in Refs. \cite{bib3}-\cite{bib5}. This mode 
is unacceptable for quantum mechanics.

Within the framework of the general relativity (GR), the coordinate 
singularity of the Schwarzschild metric (S) can be eliminated by using the 
appropriate coordinate transformations of an initial metric. Unlike the 
Schwarzschild metric, in the space-time of transformed metrics, classical 
particles can intersect the event horizon without emergence of 
singularities. Many of the researchers apply this conclusion to the quantum 
theory as well. However, it is not correct. We proved that the coordinate 
transformations of the Schwarzschild metric in the quantum mechanics of 
stationary states do not eliminate the mode of a particle ''fall'' to the 
event horizon.

As an example, it is proved by using Eddington--Finkelstein (EF) \cite{bib6, bib7}, and Painlev\`{e}--Gullstrand (PG) \cite{bib8, bib9} stationary metrics. 
This conclusion is also valid for any transformed static metrics, including 
the transition in the Schwarzschild solution to the ''tortoise'' coordinate \cite{bib10}.

The quantum mechanics in the space-time of non-stationary Lema\'{\i}tre--Finkelstein \cite{bib7, bib11},  and Kruskal- Szekeres \cite{bib12, bib13} metrics leads to 
time-dependent Hamiltonians \cite{bib14}. Study of 
stationary states of particles, with representation of wave functions in 
coordinates of these metrics in the form of $\sim \Psi \left( {{\rm {\bf 
			R}}} \right)e^{-iET}$ for the Lema\'{\i}tre-Finkelstein metric and in the 
form of $\sim \Psi \left( {{\rm {\bf u}}} \right)e^{-iE\nu }$ for the 
Kruskal--Szekeres metric, is impossible in this case.

In this paper, our study will be considered for fermions. The obtained 
results can be extended to the equations and wave functions of photons and 
spinless particles.

The paper is arranged as follows. In Sec. 2, for coherence of the 
description, we present basic features of the quantum-mechanical mode of a 
particle ''fall'' to the event horizon. In Secs. 3 and 4, we introduce the 
covariant Dirac equation and provide the bases of the theory of coordinate 
transformations of Hamiltonians for the Dirac equation in the space-time of 
(GR) metrics. These results are presented in many textbooks, monographs and 
papers (see, for instance, Refs. \cite{bib15}, \cite{bib16} as well as our papers Refs. \cite{bib14}, \cite{bib17}). In Sec. 5, 
we analyze the solutions of the Dirac equation for a Schwarzschild metric. 
In Secs. 6 -- 8, we prove the presence of the quantum-mechanical mode of 
particle ''fall'' by using stationary EF and PG metrics as well as by using the ''tortoise'' 
coordinate in the Schwarzschild static metric. In Sec. 10, we 
formulate basic results of the paper.


\section{The mode of a particle ''fall'' to the event horizon}
For all black holes, the behavior of effective potentials of the radial 
equation for Schr\"{o}dinger-type fermions in the neighborhood of event 
horizons has the form of an infinitely deep potential well \cite{bib1, bib2}
\begin{equation}
	\label{eq1}
	\left. {U_{eff} \left( r \right)} \right|_{r\to r_{\pm } } =-\frac{K_{1} 
	}{\left( {r-r_{\pm } } \right)^{2}},
\end{equation}
where $r_{\pm } $ are radii of external and internal event horizons and a 
coefficient $K_{1} >1/8$. In this case, the mode of a particle ''fall'' 
to the event horizon is implemented (see Refs. \cite{bib3} - \cite {bib5}).

The behavior of the real radial function of the Schr\"{o}dinger-type 
equation is given by\footnote{ In expression (\ref{eq2}) and in what follows, the asymptotic 
	behavior of transformation operators and wave functions is examined for 
	neighborhoods of event horizons $\left( {r>r_{+} } \right)$, $\left( 
	{r<r_{-} } \right)$.}
\begin{equation}
	\label{eq2}
	\left. {R\left( r \right)} \right|_{r\to r_{\pm } } \sim \left| {r-r_{\pm } 
	} \right|^{1/2}\sin \left( {\sqrt {K_{2} } \ln \left| 
		{\frac{r}{r_{\pm } }-1} \right|+\delta } \right),
\end{equation}
where $K_{2} =2\left( {K_{1} -1/8} \right)$. At $r\to r_{\pm } $, the radial 
functions of stationary states of discrete and continuous spectra have the 
infinite number of zeros, the discrete energy levels emerge and ''dive'' 
beyond the allowed domains of functions $R\left( r \right)$. At $r=r_{\pm } 
$, the functions of $R\left( r \right)$ do not have definite values.

In the Hamiltonian formulation, the mode of a particle ''fall'' to the 
horizon means that a Hamiltonian $H$ has non-zero deficiency indexes \cite{bib18} - \cite{bib20}.

To remove this singular mode, unacceptable for quantum mechanics, it is 
necessary to choose additional boundary conditions on event horizons. The 
self-conjugate extension of a Hermitian operator $H$ is determined by this 
choice.


\section{The Dirac equation}

In the system of units of $\hslash =c=1$ and in the signature $\left( 
{+\,\,-\,\,-\,\,-} \right)$, the Dirac equation equals
\begin{equation}
	\label{eq3}
	i\gamma^{\alpha }\left( {\frac{\partial \psi }{\partial x^{\alpha }}+\Phi 
		_{\alpha } \psi } \right)-m\psi =0,
\end{equation}
where $m$ is a fermion mass, $\Phi_{\alpha } $ are bispinor connections, 
$\psi $ is a four-component bispinor, $\gamma^{\alpha }$ are 4x4 Dirac 
matrixes with world indexes satisfying the relation of 
\begin{equation}
	\label{eq4}
	\gamma^{\alpha }\gamma^{\beta }+\gamma^{\beta }\gamma^{\alpha 
	}=2g^{\alpha \beta }E,
\end{equation}
where $g^{\alpha \beta }$ is an inverse metric tensor, $E$ is a 4x4 unity 
matrix.

In expressions (\ref{eq3}), (\ref{eq4}) and in what follows, the values designated by letters of the Greek alphabet assume the values of 0, 1, 2, 3; those designated by the 
letters of the Latin alphabet take the values of 1, 2, 3. When upper and 
lower indices are the same, the summation of appropriate summands is 
implied.

Then, along with the Dirac matrixes of $\gamma^{\alpha }$ with world 
indices, we use Dirac matrixes $\gamma^{{\underline \alpha }}$ with local indices 
satisfying the relationship of
\begin{equation}
	\label{eq5}
	\gamma^{{\underline\alpha }}\gamma^{{\underline\beta }}+\gamma^{{\underline\beta }}\gamma^{{\underline\alpha 
	}}=2\eta^{{\underline\alpha }{\underline\beta }}E,
\end{equation}
where $\eta^{{\underline\alpha }{\underline\beta }}$ corresponds to the inverse metric tensor 
of the plane Minkowski space with the signature of $\eta_{{\underline\alpha }{\underline\beta 
}} =\mbox{diag}\left[ {1,-1,-1,-1} \right].$

In Eq. (\ref{eq3}), for determination of bispinor connections $\Phi_{\alpha } 
$, it is necessary to choose a definite system of tetrad vectors $H_{{\underline\alpha 
}}^{\mu } $, satisfying the relation of $H_{{\underline\alpha }}^{\mu } H_{{\underline\beta 
}}^{\nu } g_{\mu \nu } =\eta_{{\underline\alpha }{\underline\beta }} $. The bispinor 
connections are determined by using Christoffel derivatives of tetrad 
vectors.
\begin{equation}
	\label{eq6}
	\Phi_{\alpha } =-\frac{1}{4}H_{ \mu }^{{\underline \varepsilon }} H_{\nu {\underline \varepsilon 
		};\alpha } S^{\mu \nu }=\frac{1}{4}H_{{\underline \mu }}^{\varepsilon } H_{{\underline \nu 
		}\, \varepsilon ;\alpha } S^{{\underline \mu }{\underline \nu }},
\end{equation}
where
\begin{equation}
	\label{eq7}
	\begin{array}{l}
		S^{\mu \nu }=\dfrac{1}{2}\left( {\gamma^{\mu }\gamma^{\nu }-\gamma^{\nu 
			}\gamma^{\mu }} \right), \\ [10pt]
		S^{{\underline\mu }{\underline\nu }}=\dfrac{1}{2}\left( {\gamma^{{\underline\mu }}\gamma^{{\underline\nu }}-\gamma 
			^{{\underline\nu }}\gamma^{{\underline\mu }}} \right). \\ 
	\end{array}
\end{equation}
The relation between$\gamma^{\alpha }$ and $\gamma^{{\underline \alpha }}$ is 
specified by the equality of
\begin{equation}
	\label{eq8}
	\gamma^{\alpha }=H_{{\underline\beta }}^{\alpha } \gamma^{{\underline\beta }}.
\end{equation}
For our analysis, it is convenient to use the Dirac equation in the Hamiltonian form of
\begin{equation}
	\label{eq9}
	i\frac{\partial \psi }{\partial t}=H\psi .
\end{equation}
Here $t=x^{0}$, $H$ is the Hamilton operator.

Taking into account the equation of $\gamma^{0}\gamma^{0}=g^{00}$ , we can 
derive the following expression for the Hamiltonian from Eq. (\ref{eq3})
\begin{equation}
	\label{eq10}
	H=\frac{m}{g^{00}}\gamma^{0}-\frac{1}{g^{00}}i\gamma^{0}\gamma 
	^{k}\frac{\partial }{\partial x^{k}}-i\Phi_{0} +\frac{1}{g^{00}}i\gamma 
	^{0}\gamma^{k}\Phi_{k} .
\end{equation}


\subsection{The formalism of pseudo-Hermitian quantum mechanics \cite{bib21} - \cite{bib23}}

Hamiltonians (\ref{eq10}) describing the motion of Dirac particles in arbitrary 
gravitational fields are pseudo-Hermitian \cite{bib17}.

The condition of pseudo-Hermitian character for Hamiltonians assumes the 
existence of invertible operator $\rho_{_P}$, satisfying the 
relationship of 
\begin{equation}
	\label{eq11}
	\rho_{_P} H\rho_{_P}^{-1} =H^{+}.
\end{equation}
If in this case, there exists an operator $\eta $ satisfying the condition of 
\begin{equation}
	\label{eq12}
	\rho_{_P} =\eta^{+}\eta ,
\end{equation}
then for stationary case, we obtain the following Hamiltonian in $\eta 
$-representation:
\begin{equation}
	\label{eq13}
	H_{\eta } =\eta H\eta^{-1}=H_{\eta }^{+} .
\end{equation}
Hamiltonian $H_{\eta } $ is self-conjugate with the spectrum of eigenvalues 
coinciding with the spectrum of an initial Hamiltonian $H$.

The wave-function in $\eta $-representation equals
\begin{equation}
	\label{eq14}
	\psi_{\eta } =\eta \psi ,
\end{equation}
where $\psi $ is a wave function in the Dirac equation (\ref{eq9}).

The scalar product of wave functions for pseudo-Hermitian Hamiltonians is 
written with a Parker weight operator $\rho_{_P} $ \cite{ bib14}, \cite{ bib17}, \cite{bib25}. For a wave function in $\eta $-representation, 
the scalar product has a form, typical for the Hermitian quantum mechanics 
(a plane scalar product with $\rho_{_P} =1$ )
\begin{equation}
	\label{eq15}
	\left( {\varphi_{\eta } ,\psi_{\eta } } \right)=\int {d{\rm {\bf x}}\left( 
		{\varphi_{\eta }^{+} \psi_{\eta } } \right)} .
\end{equation}
Below, for the analysis, we will use Hamiltonians and wave functions in 
$\eta $-representation.


\subsection{The system of tetrad vectors in the Schwinger gauge}

Many of the researchers use the convenient system of tetrad vectors of 
$\left\{ {\tilde{{H}}_{{\underline\alpha }}^{\mu } } \right\}$ in the Schwinger gauge \cite{bib25}. For this system,
\begin{equation}
	\label{eq16}
	\tilde{{H}}_{{\underline 0}}^{0} =\sqrt {g^{00}} ,\,\,\,\,\,\,\,\tilde{{H}}_{{\underline 0}}^{k} 
	=-{g^{k0}} / {g^{00}},\,\,\,\,\,\,\,\tilde{{H}}_{{\underline k}}^{0} =0.
\end{equation}
Any spatial tetrads, satisfying the following relations:
\begin{equation}
	\label{eq17}
	\tilde{{H}}_{{\underline k}}^{m} \tilde{{H}}_{{\underline k}}^{n} 
	=f^{mn},\,\,\,\,\,\,f^{mn}=g^{mn}+\frac{g^{0m}g^{0n}}{g^{00}},\,\,\,\,\,\,f^{mn}g_{nk} 
	=\delta_{k}^{m} 
\end{equation}
can be used as tetrad vectors of $\tilde{{H}}_{m}^{n} $.

Taking into account some freedom of choice for spatial tetrads, we can 
obtain expressions for Hamiltonians non-coincident with each other. These 
Hamiltonians are physically equivalent since they are connected by unitary 
matrixes of spatial rotations \cite{bib17}.


\section{Coordinate transformations}
At coordinate transformations (at transition to a another space-time)
\begin{equation}
	\label{eq18}
	\left\{ {x^{\alpha }} \right\}\to \left\{ {{x}'^{\alpha }} \right\}
\end{equation}
the following relationships are fulfilled:
\begin{equation}
	\label{eq19}
	{H'}_{\underline\beta }^{\alpha } =\frac{\partial {x}'^{\alpha }}{\partial x^{\mu 
	}}H_{{\underline\beta }}^{\mu } ,\,\,\,\,\,\,{\gamma }'^{\alpha }=\frac{\partial 
		{x}'^{\alpha }}{\partial x^{\beta }}\gamma^{\beta },\,\,\,\,\,\,\,{\Phi 
	}'^{\alpha }=\frac{\partial {x}'^{\beta }}{\partial x^{\alpha }}\Phi_{\beta 
	} .
\end{equation}
At transformations of (\ref{eq19}), the form of wave functions of the Dirac equation 
remains invariable except for an appropriate replacement of the variables.

In one and the same space-time, we can transfer from any system of tetrad 
vectors of $\left\{ {{H'}_{{\underline\alpha }}^{\mu } \left( x \right)} \right\}$ to 
another system of tetrad vectors of $\left\{ {H_{{\underline\alpha }}^{\mu } \left( x 
	\right)} \right\}$ by using the Lorentz transformation of $L\left( x 
\right)$. In this case,
\begin{equation}
	\label{eq20}
	H_{{\underline\alpha }}^{\mu } \left( x \right)=\Lambda_{{\underline\alpha }}^{{\underline\beta }} \left( 
	x \right){H'}_{{\underline\beta }}^{\mu } \left( x \right).
\end{equation}
The values of $\Lambda_{{\underline\alpha }}^{{\underline\beta }} $ satisfy the relations
\begin{equation}
	\label{eq21}
	\Lambda_{{\underline\alpha }}^{{\underline\mu }} \left( x \right)\Lambda_{{\underline\beta }}^{{\underline\nu }} 
	\left( x \right)\eta^{{\underline\alpha }{\underline\beta }}=\eta^{{\underline\mu }{\underline\nu }},
\end{equation}
\begin{equation}
	\label{eq22}
	\Lambda_{{\underline\alpha }}^{{\underline\mu }} \left( x \right)\Lambda_{{\underline\beta }}^{{\underline\nu }} 
	\left( x \right)\eta_{{\underline\mu }{\underline\nu }} =\eta_{{\underline\alpha }{\underline\beta }} .
\end{equation}
The matrixes of Lorentz transformation $L,L^{-1}$ are determined on the 
basis of the invariability of Dirac matrixes with local indices of $\gamma 
^{{\underline \alpha }}$ at the transformations of (\ref{eq20})
\begin{equation}
	\label{eq23}
	L\left( x \right)\gamma^{{\underline \alpha }}L^{-1}\left( x \right)=\gamma^{{\underline \beta 
	}}\Lambda_{{\underline \beta }}^{{\underline \alpha }} \left( x \right).
\end{equation}
At Lorentz transformations, Dirac currents of particles are preserved.


\section{The Schwarzschild metric}

The Schwarzschild metric in the coordinates of $\left( {t,r,\theta ,\varphi 
} \right)$ is given by
\begin{equation}
	\label{eq24}
	ds^{2}=f_{S} dt^{2}-\frac{dr^{2}}{f_{S} }-r^{2}\left( {d\theta^{2}+\sin 
		^{2}\theta d\varphi^{2}} \right),
\end{equation}
where $f_{S} =1-{r_{0}/r}$, $r_{0} ={2GM} /{c^{2}}$ is the event 
horizon (gravitational radius).


\subsection{The Dirac equation}
The nonzero tetrads in the Schwinger gauge of $\tilde{{H}}_{{\underline\alpha }}^{\mu 
} $ equal
\begin{equation}
	\label{eq25}
	\tilde{{H}}_{{\underline 0}}^{0} =1/ {\sqrt {f_{S} } 
	};\,\,\,\tilde{{H}}_{{\underline 1}}^{1} =\sqrt {f_{S} } 
	;\,\,\,\,\,\tilde{{H}}_{{\underline 2}}^{2} =1/ r;\,\,\,\,\tilde{{H}}_{{\underline 3}}^{3} =1 / {\left( {r\sin \theta } \right)}.
\end{equation}
In compliance with (\ref{eq8}), the matrixes of $\tilde{{\gamma }}^{\alpha }$ are as 
follows:
\begin{equation}
	\label{eq26}
	\tilde{{\gamma }}^{0}={\gamma^{{\underline 0}}} / {\sqrt {f_{S} 
	} },\,\,\,\,\tilde{{\gamma }}^{1}=\sqrt {f_{S} } \gamma 
	^{{\underline 1}},\,\,\,\,\tilde{{\gamma }}^{2}={\gamma^{{\underline 2}}} /r,\,\,\,\,\tilde{{\gamma }}^{3}={\gamma^{{\underline 3}}} / {\left( {r\sin \theta } \right)}.
\end{equation}
The bispinor connections of $\tilde{{\Phi }}_{\alpha } $ are as follows:
\begin{equation}
	\label{eq27}
	\begin{array}{l}
		\tilde{{\Phi }}_{0} =\dfrac{r_{0} }{4r^{2}}\gamma^{{\underline 0}}\gamma 
	^{{\underline 1}},\,\,\,\tilde{{\Phi }}_{1} =0,\,\,\,\tilde{{\Phi }}_{2} 
	=-\dfrac{1}{2}\sqrt {f_{_S} } \gamma^{{\underline 1}}\gamma^{{\underline 2}}, \\ [5pt]
	\tilde{{\Phi 
	}}_{3} =-\dfrac{1}{2}\cos \theta \gamma^{{\underline 2}}\gamma^{{\underline 3}}+\dfrac{1}{2}\sqrt 	{f_{S} } \sin \theta \gamma^{{\underline 3}}\gamma^{{\underline 1}}.
\end{array}
\end{equation}
Taking into account (\ref{eq25}) - (\ref{eq27}), the Dirac Hamiltonian of (\ref{eq10}) is given by
\begin{equation}
	\label{eq28}
	\begin{array}{l}
		\tilde{{H}}_{S} =\sqrt {f_{S} } m\gamma^{{\underline 0}}-i\sqrt {f_{S} } \gamma 
		^{{\underline 0}}\left\{ {\gamma^{{\underline 1}}\sqrt {f_{S} } \left( {\dfrac{\partial }{\partial 
					r}+\dfrac{1}{r}} \right)+\gamma^{{\underline 2}}\dfrac{1}{r}\left( {\dfrac{\partial 
				}{\partial \theta }+\dfrac{1}{2}\mbox{ctg}\theta } \right)} \right.+ \\ [10pt]
		\left. {+\gamma^{{\underline 3}}\dfrac{1}{r\sin \theta }\dfrac{\partial }{\partial 
				\varphi }} \right\} -i\dfrac{r_{0} }{4r^{2}}\gamma^{{\underline 0}}\gamma^{{\underline 1}}. \\ 
	\end{array}
\end{equation}
Hamiltonian (\ref{eq28}) is pseudo-Hermitian with the weight operator $\rho_{_P} 
=f_{S}^{-1/2}$ \cite{bib14}, \cite{bib17}, \cite{bib25}.

The self-conjugate Hamiltonian with a plane scalar product of wave functions has the following form (see Refs. \cite{bib14} and \cite{bib17}):
\begin{equation}
	\label{eq29}
	\begin{array}{l}
		H_{\eta_{_S} } =H_{\eta_{_S} }^{+} =\eta_{_S} \tilde{{H}}_{S} \eta 
		_{_S}^{-1} =\sqrt {f_{S} } m\gamma^{{\underline 0}}-i\sqrt {f_{S} } \gamma 
		^{{\underline 0}}\left\{ {\gamma^{{\underline 1}}\sqrt {f_{S} } \left( {\dfrac{\partial }{\partial 
					r}+\dfrac{1}{r}} \right)} \right. + \\ [10pt]
			\left. {+ \gamma^{{\underline 2}}\dfrac{1}{r}\left( {\dfrac{\partial 
				}{\partial \theta }+\dfrac{1}{2}\mbox{ctg}\theta } \right)+ \gamma^{{\underline 3}}\dfrac{1}{r\sin \theta }\dfrac{\partial }{\partial 
				\varphi }} \right\} -i\dfrac{r_{0} }{2r^{2}}\gamma^{{\underline 0}}\gamma^{{\underline 1}}, \\ 
	\end{array}
\end{equation}
where $\eta_{_S} =f_{S}^{-1/4}$.

For separation of variables, let us present a bispinor $\psi_{\eta_{_S} } 
\left( {t,r,\theta ,\varphi } \right)$ as
\begin{equation}
	\label{eq30}
	\psi_{\eta_{_S} } \left( {t,r,\theta ,\varphi } \right)=\left( 
	{\begin{array}{l}
			\,\,\,\,\,F\left( r \right)\xi \left( \theta \right) \\ 
			-iG\left( r \right)\sigma^{3}\xi \left( \theta \right)\,\,\,\, \\ 
	\end{array}} \right)e^{-iEt}e^{im_{\varphi } \varphi }.
\end{equation}
Hereinafter, $\sigma^{3}$ is a two-dimensional Pauli matrix.

Spinor $\xi \left( \theta \right)$ represents spherical harmonics of spin 
one-half. $E,m$ are energy and mass of the Dirac particle, $m_{\varphi } 
=-j,-j+1,...j-1$, $j$ is an azimuth component of a total momentum $j$, 
$\kappa $ is a quantum number of the Dirac equation:
\[
\kappa =\mp 1,\mp 2...=\left\{ {\begin{array}{l}
		-\left( {l+1} \right),\,\,\,j=l+\frac{1}{2} \\ 
		\,\,\,\,\,\,\,\,\,l,\,\,\,\,\,\,\,\,j=l-\frac{1}{2}, \\ 
\end{array}} \right.
\]
where $j,l$ are quantum numbers of the total and orbital momenta of a Dirac 
particle.

For separation of variables , it is convenient to perform the following 
equivalent substitution of matrixes:
\begin{equation}
	\label{eq31}
	\gamma^{\underline 1} \to \gamma^{\underline 3},\,\,\,\,\,	\gamma^{\underline 3} \to \gamma^{\underline 2},\,\,\,\,\,
	\gamma^{\underline 2} \to \gamma^{\underline 1}.
\end{equation}
After separation of the variables, the equations for radial functions have 
the following form:
\begin{equation}
	\label{eq32}
	\begin{array}{l}
		f_{S} \dfrac{dF}{dr}+\left( {\dfrac{1+\kappa \sqrt {f_{S} } }{r}-\dfrac{r_{0} 
			}{2r^{2}}} \right)F-\left( {E+m\sqrt {f_{S} } } \right)G=0, \\ [10pt] 
		f_{S} \dfrac{dG}{dr}+\left( {\dfrac{1-\kappa \sqrt {f_{S} } }{r}-\dfrac{r_{0} 
			}{2r^{2}}} \right)G+\left( {E-\sqrt {f_{S} } } \right)F=0. \\ 
	\end{array}
\end{equation}
If at $r\to r_{0} $, we present the functions of $F\left( \rho 
\right),G\left( \rho \right)$, as follows:
\begin{equation}
	\label{eq33}
	\left. F \right|_{r\to r_{0} } =\left( {r-r_{0} } 
	\right)^{p}\sum\limits_{k=0}^\infty {f_{k} \left( {r-r_{0} } \right)^{k}} 
	,\,\,\,\left. G \right|_{r\to r_{0} } =\left( {r-r_{0} } 
	\right)^{p}\sum\limits_{k=0}^\infty {g_{k} \left( {r-r_{0} } \right)^{k}} ,
\end{equation}
then, the indicial equation for system (\ref{eq32}) leads to the two solutions with 
identical sense:
\begin{equation}
	\label{eq34}
	\left( p \right)_{1,2} =-\frac{1}{2}\pm ir_{0} E.
\end{equation}
\begin{equation}
	\label{eq35}
	\left. {F_{1} } \right|_{r\to r_{0} } , \,\,\left. {G_{1} } \right|_{r\to r_{0} } 
	\sim \frac{1}{\left( {r-r_{0} } \right)^{1/2}}\exp \left\{ {ir_{0} E\ln \left( 
		{\frac{r}{r_{0} }-1} \right)} \right\},
\end{equation}
\begin{equation}
	\label{eq36}
	\left. {F_{2} } \right|_{r\to r_{0} } , \,\,\left. {G_{2} } \right|_{r\to r_{0} } 
	\sim \frac{1}{\left( {r-r_{0} } \right)^{1 / 2}}\exp \left\{ {-ir_{0} E\ln \left( 
		{\frac{r}{r_{0} }-1} \right)} \right\}.
\end{equation}
In this case, solutions (\ref{eq32}) can be represented by real functions \cite{bib3, bib5}. Taking into account (\ref{eq34}), we can write down the asymptotics of the solutions at $r\to r_{0} $ as
\begin{equation}
	\label{eq37}
	\begin{array}{l}
		\left. F \right|_{r\to r_{0} } =\dfrac{L}{\sqrt {r-r_{0} } }\sin \left( 
		{r_{0} E\ln \left( {\dfrac{r}{r_{0} }-1} \right)+\delta } \right), \\ [10pt]
		\left. G \right|_{r\to r_{0} } =\dfrac{L}{\sqrt {r-r_{0} } }\cos \left( 
		{r_{0} E\ln \left( {\dfrac{r}{r_{0} }-1} \right)+\delta } \right), \\ 
	\end{array}
\end{equation}
where $L,\delta $ are integration constants. In some of the scattering 
problems, one can use a complex phase $\delta $ \cite{bib5}.

In expressions (\ref{eq35}) - (\ref{eq37}), the functions of $F\left( r \right),G\left( r \right)$ are square-nonintegrable at $r\to r_{0} $. The oscillating form of the functions of $F\left( r \right),G\left( r \right)$ for $E\ne 0$ 
testifies to implementation of the mode of a particle ''fall'' to event 
horizons (see Eq. (\ref{eq2}) in Sec. 2).

The solutions of the indicial equation for the system of equations for 
radial functions of Hamiltonian (\ref{eq28}) are as follows: 
\begin{equation}
	\label{eq38}
	\left( {p_{S} } \right)_{1,2} =-\frac{1}{4}\pm ir_{0} E.
\end{equation}
The asymptotics of the radial functions at $r\to r_{0} $ are given by
\begin{equation}
	\label{eq39}
	\left. {\left( {F_{S} } \right)_{1} } \right|_{r\to r_{0} } , \,\,\left. {\left( 
		{G_{S} } \right)_{1} } \right|_{r\to r_{0} } \sim \frac{1}{\left( {r-r_{0} } 
		\right)^{1 /4}}\exp \left\{ {ir_{0} E\ln \left( 
		{\frac{r}{r_{0} }-1} \right)} \right\},
\end{equation}
\begin{equation}
	\label{eq40}
	\left. {\left( {F_{S} } \right)_{2} } \right|_{r\to r_{0} } , \,\,\left. {\left( 
		{G_{S} } \right)_{2} } \right|_{r\to r_{0} } \sim \frac{1}{\left( {r-r_{0} } 
		\right)^{1 / 4}}\exp \left\{ {-ir_{0} E\ln \left( 
		{\frac{r}{r_{0} }-1} \right)} \right\}.
\end{equation}
Functions $F_{S} \left( r \right),G_{S} \left( r \right)$ are 
square-nonintegrable due to the presence of Parker weight operator $\rho_{_P} 
=f_{S}^{-1/ 2} $ in the scalar product \cite{bib14}, \cite{bib17}, \cite{bib25}. In this case, the mode of a particle ''fall'' to the event horizon of $r_{0} $ is also present. 

Below, the problem of the possibility to eliminate the mode of a particle 
''fall'' by means of the transformations of the coordinates of the 
Schwarzschild's metric is analyzed.


\section{The stationary Eddington-Finkelstein metric}
\label{sec:mylabel5}
The coordinate transformation of the Schwarzschild metric
\begin{equation}
	\label{eq41}
	\left( {t,r,\theta ,\varphi } \right)\to \left( {T,r,\theta ,\varphi } 
	\right)
\end{equation}
has the form of 
\begin{equation}
	\label{eq42}
	dT=dt+\frac{r_{0} }{r}\frac{dr}{f_{S} }.
\end{equation}
Then, the EF metric is written as
\begin{equation}
	\label{eq43}
	ds^{2}=f_{S} dT^{2}-2\frac{r_{0} }{r}dTdr-\left( {1+\frac{r_{0} }{r}} 
	\right)dr^{2}-r^{2}\left( {d\theta^{2}+\sin^{2}\theta d\varphi^{2}} 
	\right).
\end{equation}
From Hamiltonian $\tilde{{H}}_{S} $ in the Schwarzschild space--time (\ref{eq28}), we 
will obtain Hamiltonian $H_{EF} $ in the EF space--time, by using relation 
(\ref{eq19}) and Eqs. (\ref{eq25}) - (\ref{eq27}).

As the result,
\begin{equation}
	\label{eq44}
	\begin{array}{l}
	\left( {H_{EF} } \right)_{{\underline 0}}^{0} =\dfrac{1}{\sqrt {f_{S} } },\,\,\,\,\left( 
	{H_{EF} } \right)_{{\underline 1}}^{0} =\dfrac{{r_{0} }/  r}{\sqrt {f_{S} } 
	},\,\,\,\,\left( {H_{EF} } \right)_{{\underline 1}}^{1} =\sqrt {f_{S} } , \\ [10pt]
\left( 	{H_{EF} } \right)_{{\underline 2}}^{2} =\dfrac{1}{r},\,\,\,\,\,\,\,\left( {H_{EF} } 
	\right)_{{\underline 3}}^{3} =\dfrac{1}{r\sin \theta },
	\end{array}
\end{equation}
\begin{equation}
	\label{eq45}
	\gamma_{EF}^{0} =\frac{1}{\sqrt {f_{S} } }\gamma^{{\underline 0}} +\frac{r_{0} /r}{\sqrt {f_{S} } }\gamma^{{\underline 1}} ,\,\,\,\,\gamma_{EF}^{1} =\sqrt {f_{S} 
	} \gamma^{{\underline 1}} ,\,\,\,\gamma_{EF}^{2} =\frac{1}{r}\gamma^{{\underline 2}} 
	,\,\,\,\,\,\,\,\gamma_{EF}^{3} =\frac{1}{r\sin \theta }\gamma^{{\underline 3}} ,
\end{equation}
\begin{equation}
	\label{eq46}
	\begin{array}{l}
		\left( {\Phi_{EF} } \right)_{0} =\dfrac{r_{0} }{4r^{2}}\gamma^{{\underline0}} 
		\gamma^{{\underline1}} ,\,\,\,\,\,\,\,\left( {\Phi_{EF} } \right)_{1} 
		=\dfrac{r_{0} }{r}\dfrac{1}{f_{S} }\dfrac{r_{0} }{4r^{2}}\gamma^{{\underline 0}} 
		\gamma^{{\underline 1}} , \\ [10pt]
		\left( {\Phi_{EF} } \right)_{2} =-\dfrac{1}{2}\sqrt {f_{S} } \gamma 
		^{{\underline 1}} \gamma^{{\underline 2}} ,\,\,\,\,\,\,\,\left( {\Phi_{EF} } \right)_{3} 
		=-\dfrac{1}{2}\cos \theta \,\gamma^{{\underline 2}} \gamma^{{\underline 3}} 
		+\dfrac{1}{2}\sin \theta \gamma^{{\underline 3}} \gamma^{{\underline 1}} . \\ 
	\end{array}
\end{equation}
For the EF metric (see Eq. (\ref{eq43})), the inverse metric tensor is
\begin{equation}
	\label{eq47}
	g^{00}=1+\frac{r_{0} }{r}.
\end{equation}
By substituting (\ref{eq44}) - (\ref{eq47}) to Eq. (\ref{eq10}), we obtain the Hamiltonian in the EF space--time
\begin{equation}
	\label{eq48}
	H_{EF} =\frac{1-\left( {r_{0}/ r} \right)\gamma^{{\underline 0}} \gamma 
		^{{\underline 1}} }{1-{r_{0}^{2} }/ {r^{2}}}\tilde{{H}}_{S} .
\end{equation}
Inverse equality is
\begin{equation}
	\label{eq49}
	\tilde{{H}}_{S} =\left( {1+\frac{r_{0} }{r}\gamma^{{\underline 0}} \gamma^{{\underline 1}} 
	} \right)H_{EF} .
\end{equation}
Let us demonstrate that Hamiltonians $H_{EF} $ and $\tilde{{H}}_{S} $ are 
connected with each other through the unitary transformation.

At transformations (\ref{eq19}), the form of wave functions does not vary except for 
substitution of the variables. Let us denote
\begin{equation}
	\label{eq50}
	\varphi_{_{EF}} \left( r \right)=\int {\frac{r_{0} }{r}\frac{dr}{f_{S} }} .
\end{equation}
It follows from (\ref{eq42}), that $t=T-\varphi_{_{EF}} \left( r \right)=T-\int 
{\dfrac{r_{0} }{r}\dfrac{dr}{f_{S} }} =T-\left( {r_{0} \mbox{ln}\left( 
	{\frac{r}{r_{0} }-1} \right)+\mbox{const}} \right)$. Since
\begin{equation}
	\label{eq51}
	\tilde{{\psi }}_{S} \left( {{\rm {\bf r}},t} \right)=\psi_{_{EF}} \left( {{\rm 
			{\bf r}},T} \right),
\end{equation}
then
\begin{equation}
	\label{eq52}
	\tilde{{\psi }}_{S} \left( {{\rm {\bf r}}} \right)e^{-iEt}=\tilde{{\psi 
	}}_{S} \left( {{\rm {\bf r}}} \right)e^{iE\varphi_{_{EF}} \left( r 
		\right)}e^{-iET}=\psi_{_{EF}} \left( {{\rm {\bf r}}} \right)e^{-iET}.
\end{equation}
The equation for stationary states in the Schwarzschild space--time equals 
\begin{equation}
	\label{eq53}
	\tilde{{H}}_{S} \tilde{{\psi }}_{S} \left( {{\rm {\bf r}}} 
	\right)=E\tilde{{\psi }}_{S} \left( {{\rm {\bf r}}} \right).
\end{equation}
It follows from (\ref{eq52}), that $\tilde{{\psi }}_{S} \left( {{\rm {\bf r}}} 
\right)=e^{-iE\varphi_{EF} \left( r \right)}\psi_{EF} \left( {{\rm {\bf 
			r}}} \right)$, and we obtain that
\begin{equation}
	\label{eq54}
	e^{iE\varphi_{_{EF}} \left( r \right)}\tilde{{H}}_{S} e^{-iE\varphi_{_{EF}} 
		\left( r \right)}=E\psi_{EF} \left( {{\rm {\bf r}}} \right).
\end{equation}
Using the explicit form of $\tilde{{H}}_{S} $ in (\ref{eq28}) and equality (\ref{eq49}), we obtain that
\begin{equation}
	\label{eq55}
	\left( {1+\frac{r_{0} }{r}\gamma^{{0}} \gamma^{{1}} } \right)\left[ 
	{H_{EF} \psi_{EF} \left( {{\rm {\bf r}}} \right)-E\psi_{EF} \left( {{\rm 
				{\bf r}}} \right)} \right]=0.
\end{equation}
By multiplying (\ref{eq55}) on the left by matrix ${\left( {1-\frac{r_{0} }{r}\gamma 
		^{{\underline 0}} \gamma^{{\underline 1}} } \right)} \mathord{\left/ {\vphantom {{\left( 
				{1-\frac{r_{0} }{r}\gamma^{{0}} \gamma^{{1}} } \right)} {\left( 
				{1-{r_{0}^{2} }/ {r^{2}}} \right)}}} \right. 
	\kern-\nulldelimiterspace} {\left( {1-{r_{0}^{2} } /{r^{2}}} \right)}$, we will obtain an equation for stationary states in the 
EF space--time:
\begin{equation}
	\label{eq56}
	H_{EF} \psi_{EF} \left( {{\rm {\bf r}}} \right)=E\psi_{EF} \left( {{\rm 
			{\bf r}}} \right).
\end{equation}
In this case, in compliance with (\ref{eq54}) and (\ref{eq52}), the following relations are 
valid:
\begin{equation}
	\label{eq57}
	H_{EF} =e^{iE\varphi_{_{EF}} }\tilde{{H}}_{S} e^{-iE\varphi_{_{EF}} },
\end{equation}
\begin{equation}
	\label{eq58}
	\psi_{EF} \left( {{\rm {\bf r}}} \right)=\tilde{{\psi }}_{S} \left( {{\rm 
			{\bf r}}} \right)e^{iE\varphi_{_{EF}} }.
\end{equation}
Thus, it is demonstrated that Hamiltonians $H_{EF} $ and $\tilde{{H}}_{S} $ 
are connected with each other through the unitary transformation:
\begin{equation}
	\label{eq59}
	U_{EF} =e^{iE\varphi_{_{EF}} \left( r \right)}.
\end{equation}
It follows from (\ref{eq58}) that the asymptotics of radial functions in the 
EF space-time can be obtained by multiplication of 
asymptotics (\ref{eq39}) and (\ref{eq40}) by unitary operator (\ref{eq59}). Then,
\begin{equation}
	\label{eq60}
	\left. {\left( {F_{EF} } \right)_{1} } \right|_{r\to r_{0} } , \,\,\left. {\left( 
		{G_{EF} } \right)_{1} } \right|_{r\to r_{0} } \sim \frac{1}{\left( {r-r_{0} 
		} \right)^{1 / 4}}\exp \left\{ {i2r_{0} E\ln \left( 
		{\frac{r}{r_{0} }-1} \right)} \right\},
\end{equation}
\begin{equation}
	\label{eq61}
	\left. {\left( {F_{EF} } \right)_{2} } \right|_{r\to r_{0} } , \,\,\left. {\left( 
		{G_{EF} } \right)_{2} } \right|_{r\to r_{0} } \sim \frac{1}{\left( {r-r_{0} 
		} \right)^{1 / 4}}.
\end{equation}
The asymptotics (\ref{eq60}) and (\ref{eq61}) can be obtained by using another way. 

We can separate variables in Eq. (\ref{eq56}) by representing the wave function 
in the form of (\ref{eq30}). The equations for radial functions of $F_{EF} \left( r 
\right),G_{EF} \left( r \right)$ are given by
\begin{equation}
	\label{eq62}
	\begin{array}{l}
		f_{S} \dfrac{dF_{EF} }{dr}+\left( {\dfrac{1+\kappa \sqrt {f_{S} } 
			}{r}-\dfrac{3}{4}\dfrac{r_{0} }{r^{2}}} \right)F_{EF} -i\dfrac{r_{0} 
		}{r}EF_{EF} -\left( {E+\sqrt {f_{S} } m} \right)G_{EF} =0, \\ [10pt]
		f_{S} \dfrac{dG_{EF} }{dr}+\left( {\dfrac{1-\kappa \sqrt {f_{S} } 
			}{r}-\dfrac{3}{4}\dfrac{r_{0} }{r^{2}}} \right)G_{EF} -i\dfrac{r_{0} 
		}{r}EG_{EF} +\left( {E-\sqrt {f_{S} } m} \right)F_{EF} =0. \\ 
	\end{array}
\end{equation}
To determine the asymptotics of radial functions at $r\to r_{0} $, let us 
present functions $\left. {\left( {F_{EF} } \right)} \right|_{r\to r_{0} } 
,\left. {\left( {G_{EF} } \right)} \right|_{r\to r_{0} } $ in the form of 
(\ref{eq33}). Let us write the indicial equation for system (\ref{eq62}) as an equation for the determinant:
\begin{equation}
	\label{eq63}
	\begin{vmatrix}
				{\left( {p+\dfrac{1}{4}} \right)\dfrac{1}{r_{0} }-iE} \hfill &\,\,\,\,\,\,\,\,\,\,\,\,\,\,\,{-E} \hfill \\ 
				\,\,\,\,\,\,\,\,\,\,\,\,\,\,\,\,\,\,\,\,\,\,E \hfill & {\left( {p+\dfrac{1}{4}} \right)\dfrac{1}{r_{0} }-iE} \hfill \\
	\end{vmatrix} =0.
\end{equation}
The solutions for Eq. (\ref{eq63}) are given by
\begin{equation}
	\label{eq64}
	\begin{array}{l}
		p_{1} =-\dfrac{1}{4}+2ir_{0} E, \\ [5pt]
		p_{2} =-\dfrac{1}{4}. \\ 
	\end{array}
\end{equation}
Equalities (\ref{eq64}) show that the required asymptotics coincide with asymptotics 
(\ref{eq60}), (\ref{eq61}).

The Parker operator in the (EF) space-time with the set of tetrads (\ref{eq44}) is 
\begin{equation}
	\label{eq65}
	\left( {\rho_{_P} } \right)_{EF} =\gamma^{{\underline 0}}\gamma_{EF}^{0} 
	=\frac{1}{\sqrt {f_{S} } }\left( {1+\frac{r_{0} }{r}\gamma^{{\underline 0}}\gamma 
		^{{\underline 1}}} \right).
\end{equation}
For both asymptotics (\ref{eq60}), (\ref{eq61}), the scalar product in the neighborhood of the event horizon diverge lolgarithmically.

We can formally determine the form of the operator of transition to a 
self-conjugate Hamiltonian in the EF space-time with a plane scalar 
product Eq. (see (\ref{eq15}))
\begin{equation}
	\label{eq66}
	\begin{array}{l}
		\left( {\rho_{_P} } \right)_{EF} =\eta_{_{EF}}^{+} \eta_{_{EF}} , \\ 
		\eta_{_{EF}} =\dfrac{1}{f_{S}^{1 / 4} }\left( {1+\dfrac{r_{0} }{r}\gamma^{\underline{0}}\gamma 
			^{\underline{1}}} \right)^{1 /2}, \\ 
	\end{array}
\end{equation}
where $\eta_{_{EF}}^{+} =\eta_{_{EF}} $. However, the application of a matrix 
operator $\eta_{_{EF}} $ is difficult in practice.


\subsection{The Lorentz transformation}

To obtain an operator $\eta_{_{EF}} $ in an acceptable matrix-free form, let 
us transfer from the system of tetrad vectors of $\left\{ {\left( {H_{EF} } 
	\right)_{{\underline \alpha }}^{\mu } } \right\}$ to tetrad vectors in the Schwinger 
gauge of $\left\{ {\left( {\tilde{{H}}_{EF} } \right)_{{\underline \alpha }}^{\mu } } 
\right\}$ by using Lorentz transformation of (\ref{eq20}) - (\ref{eq23}). In this system, the tetrads of $\tilde{{H}}_{{\underline k}}^{0} $ are equal to zero and an operator 
$\eta_{_{EF}} $ is free from matrixes $\gamma^{{\underline 0}},\gamma^{{\underline 1}}$. The 
nonzero tetrads of $\left( {\tilde{{H}}_{EF} } \right)_{{\underline \alpha }}^{\mu } $ 
in the Schwinger gauge are determined by the expressions of
\begin{equation}
	\label{eq67}
	\begin{array}{l}
		\left( {\tilde{{H}}_{EF} } \right)_{{\underline 0}}^{0} =\sqrt {1+\dfrac{r_{0} }{r}} 
		,\,\,\,\,\,\,\left( {\tilde{{H}}_{EF} } \right)_{{\underline 0}}^{1} =-\dfrac{{r_{0} } 
			/r}{\sqrt {1+{r_{0} } / r} },\,\,\,\,\,\left( {\tilde{{H}}_{EF} } 
		\right)_{{\underline 1}}^{1} =\dfrac{1}{\sqrt {1+{r_{0} } / r} }, \\ [10pt]
		\,\left( {\tilde{{H}}_{EF} } \right)_{{\underline 2}}^{2} 
		=\dfrac{1}{r},\,\,\,\,\,\,\,\left( {H_{EF} } \right)_{{\underline 3}}^{3} 
		=\dfrac{1}{r\sin \theta }. \\ 
	\end{array}
\end{equation}
In considered case, the nonzero values of $\Lambda_{{\underline \alpha }}^{{\underline \beta }} 
\left( r \right)$ in (\ref{eq20}) are equal to:
\begin{equation}
	\label{eq68}
	\Lambda_{{\underline 0}}^{{\underline 0}} =\Lambda_{{\underline 1}}^{{\underline 1}} =\frac{1}{\sqrt {f_{S} } \sqrt 
		{1+{r_{0} } /r} },\,\,\,\,\,\Lambda_{{\underline 0}}^{{\underline 1}} =\Lambda 
	_{{\underline 1}}^{{\underline 0}} =-\frac{{r_{0} } / r}{\sqrt {f_{S} } \sqrt {1+{r_{0} } /r} }.
\end{equation}
Let us determine the form of the Lorentz transformation matrixes from 
Eqs. (\ref{eq23}) with the values of $\Lambda_{{\underline \alpha }}^{{\underline \beta }} \left( x 
\right)$ from (\ref{eq68})
\begin{equation}
	\label{eq69}
	L\gamma^{{\underline 0}}L^{-1}=\frac{1}{\sqrt {1-{r_{0} }/ r} \sqrt {1+{r_{0} } 
			/r} }\left( {\gamma^{{\underline 0}}-\frac{r_{0} }{r}\gamma^{{\underline 1}} } \right),\,
\end{equation}
\begin{equation}
	\label{eq70}
	L\gamma^{{\underline 1}}L^{-1}=\frac{1}{\sqrt {1-{r_{0} } / r} \sqrt {1+{r_{0} } 
			/r} }\left( {-\frac{r_{0} }{r}\gamma^{{\underline 0}}+\gamma^{{\underline 1}} } \right).
\end{equation}
It is well--known that Lorentz transformations can unambiguously be presented 
as a product of either a boost (a Hermitian factor) by a spatial rotation 
matrix $R$ (a unitary factor), or, vice versa, as a product of a matrix $R$ 
by a boost. For our case, it is sufficient to use a unity matrix $R=1$. 

The following forms of $L$ and $L^{-1}$ are reasonable:
\begin{equation}
	\label{eq71}
	\begin{array}{l}
		L=\mbox{exp}\left( {\dfrac{\theta }{2}\gamma^{{\underline 0}}\gamma^{{\underline 1}} } 
		\right)=\mbox{ch}\dfrac{\theta }{2}+\mbox{sh}\dfrac{\theta }{2}\gamma 
		^{{\underline 0}}\gamma^{{\underline 1}} =\dfrac{1+\left( {B/ A} \right)\gamma^{{\underline 0}}\gamma^{{\underline 1}} 
		}{\sqrt {1-{B^{2}} / {A^{2}}} },\,\,\,\,\, \\ [10pt]
		L^{-1}=\mbox{exp}\left( {-\dfrac{\theta }{2}\gamma^{{\underline 0}}\gamma^{{\underline 1}} } 
		\right)=\mbox{ch}\dfrac{\theta }{2}-\mbox{sh}\dfrac{\theta }{2}\gamma 
		^{{\underline 0}}\gamma^{{\underline 1}} =\dfrac{1-\left( {B / A} \right)\gamma^{{\underline 0}}\gamma^{{\underline 1}} 
		}{\sqrt {1-{B^{2}} / {A^{2}}} }. \\ 
	\end{array}
\end{equation}
In (\ref{eq71}), the matrix $L$ represents a transformation of hyperbolic rotation 
(i.e. boost) by an angle $\theta $.

Substituting (\ref{eq71}) in Eqs. (\ref{eq69}) and (\ref{eq70}), we obtain that
\begin{equation}
	\label{eq72}
	\frac{B}{A}=\frac{\sqrt {1+{r_{0} } /r} -\sqrt {1-{r_{0} } / r} }{\sqrt 
		{1+{r_{0} } / r} +\sqrt {1-{r_{0} } / r} }.
\end{equation}
Then,
\begin{equation}
	\label{eq73}
	L\left( r \right)=\frac{\sqrt {1+{r_{0} } / r} +\sqrt {1-{r_{0} } 
			/r} +\left( {\sqrt {1+{r_{0} } / r} -\sqrt {1-{r_{0} } /r} } 
		\right)\gamma^{{\underline 0}}\gamma^{{\underline 1}} }{2\sqrt {\sqrt {1+{r_{0} } 
			/r} \sqrt {1-{r_{0} } / r} } },
\end{equation}
\begin{equation}
	\label{eq74}
	L^{-1}\left( r \right)=\frac{\sqrt {1+{r_{0} } / r} +\sqrt {1-{r_{0} } 
			/r} +\left( {\sqrt {1+{r_{0} } /r} -\sqrt {1-{r_{0} } / r} } 
		\right)\gamma^{{\underline 1}} \gamma^{{\underline 0}}}{2\sqrt {\sqrt {1+{r_{0} } 
				/r} \sqrt {1-{r_{0} } /r} } }.
\end{equation}
After Lorenz transformation of (\ref{eq73}) and (\ref{eq74}) , Hamiltonian (\ref{eq57}) is transformed to the Hamiltonian with tetrads in Schwinger gauge \cite{bib14}:
\begin{equation}
	\label{eq75}
	\begin{array}{l}
		\tilde{{H}}_{EF} =L\left( r \right)H_{EF} L^{-1}\left( r \right)= \\ 
		=\dfrac{m}{\sqrt {1+{r_{0} } /r} }\gamma^{{\underline 0}}-i\gamma^{{\underline 0}}\gamma 
		^{{\underline 1}}\dfrac{1}{1+{r_{0} } / r}\left( {\dfrac{\partial }{\partial 
				r}+\dfrac{1}{r}+\dfrac{r_{0} }{4r^{2}}\dfrac{1}{1+{r_{0} } / r}} 
		\right)- \\ [10pt]
		- i\gamma^{{\underline 0}}\gamma^{{\underline 2}}\dfrac{1}{\sqrt {1+{r_{0} } 
				/r} }\dfrac{1}{r}\left( {\dfrac{\partial }{\partial \theta 
			}+\dfrac{1}{2}\mbox{ctg}\theta } \right)-i\gamma^{{\underline 0}}\gamma^{{\underline 3}}\dfrac{1}{\sqrt {1+{r_{0} } / r} 
		}\dfrac{1}{r\,\mbox{sin}\theta }\dfrac{\partial }{\partial \varphi 
		}+ \\ [10pt]
	+i\dfrac{r_{0} }{r}\dfrac{1}{1+{r_{0} } / r}\left( {\dfrac{\partial }{\partial 
				r}+\dfrac{1}{r}-\dfrac{1}{4r\left( {1+{r_{0} } / r} \right)}-\dfrac{1}{4r}} 
		\right). \\ 
	\end{array}
\end{equation}
In this case,
\begin{equation}
	\label{eq76}
	\tilde{{\psi }}_{EF} =L\left( r \right)\psi_{EF} .
\end{equation}
According to (\ref{eq73}), the asymptotics of $L\left( r \right)$ at $r\to r_{0} $ 
are
\begin{equation}
	\label{eq77}
	\left. L \right|_{r\to r_{0} } =\frac{r_{0}^{1 /4} \left( {1+\gamma^{{\underline 0}}\gamma 
			^{{\underline 1}} } \right)}{2^{3 /4}\left( {r-r_{0} } \right)^{1/ 4}}.
\end{equation}
In compliance with (\ref{eq76}) and (\ref{eq77}), after the similarity transformation $L\left( r \right)$, the asymptotics of radial functions (\ref{eq60}) and (\ref{eq61}) are transformed to the following form:
\begin{equation}
	\label{eq78}
	\left. {\left( {\tilde{{F}}_{EF} } \right)_{1} } \right|_{r\to r_{0} } 
	, \,\,\left. {\left( {\tilde{{G}}_{EF} } \right)_{1} } \right|_{r\to r_{0} } \sim 
	\frac{1}{\left( {r-r_{0} } \right)^{1 / 2}}\exp \left\{ {i2r_{0} E\ln \left( 
		{\frac{r}{r_{0} }-1} \right)} \right\},
\end{equation}
\begin{equation}
	\label{eq79}
	\left. {\left( {\tilde{{F}}_{EF} } \right)_{2} } \right|_{r\to r_{0} } 
	, \,\,\left. {\left( {\tilde{{G}}_{EF} } \right)_{2} } \right|_{r\to r_{0} } \sim 
	\frac{1}{\left( {r-r_{0} } \right)^{1 / 2}}.
\end{equation}
Now, let us obtain asymptotics of radial functions by using an another 
technique. If we present the wave function of $\tilde{{\psi }}_{EF} \left( 
{{\rm {\bf r}},T} \right)$ in a form similar to Eq. (\ref{eq30}), then in the 
equation of 
\begin{equation}
	\label{eq80}
	\tilde{{H}}_{EF} \tilde{{\psi }}_{EF} =E\tilde{{\psi }}_{EF} 
\end{equation}
we can perform separation of the variables. So,
\begin{equation}
	\label{eq81}
	\tilde{{\psi }}_{EF} \left( {r,\theta ,\varphi ,T} \right)=\left( 
	{\begin{array}{l}
			\,\,\,\,\,\tilde{{F}}_{EF} \left( r \right)\xi \left( \theta \right) \\ 
			-i\tilde{{G}}_{EF} \left( r \right)\sigma^{3}\xi \left( \theta 
			\right)\,\,\,\, \\ 
	\end{array}} \right)e^{-iET+im_{\varphi } \varphi }.
\end{equation}
For convenience, while separating the variables, the equivalent substitution 
of matrixes $\gamma^{{\underline i}}$ is carried out (see Eq. (\ref{eq31})). The system of 
equations for radial wave functions of $\tilde{{F}}_{EF} \left( r 
\right),\tilde{{G}}_{EF} \left( r \right)$ is given by
\begin{equation}
	\label{eq82}
	\begin{array}{l}
		\left( {1-\dfrac{r_{0} }{r}} \right)\dfrac{d\tilde{{F}}_{EF} }{dr}+\left( 
		{\dfrac{1-{r_{0} } / r+\kappa / {\sqrt 
					{1+{r_{0} } / r} }}{r}+\dfrac{r_{0} }{4r^{2}}} 
		\right)\tilde{{F}}_{EF} - \\ [10pt]
		- i\dfrac{r_{0} }{r}\left( {E-\dfrac{m}{\sqrt {1+{r_{0} 
					} / r} }} \right)\tilde{{F}}_{EF} + i\left( {-\dfrac{1}{1+{r_{0} } / r}\dfrac{r_{0} }{2r^{2}}+\dfrac{r_{0} 
			}{r^{2}}\dfrac{\kappa }{\sqrt {1+{r_{0} } / r} }} \right)\tilde{{G}}_{EF} - \\ [10pt]
		- \left( {E+\dfrac{m}{\sqrt {1+{r_{0} } / r} }} \right)\tilde{{G}}_{EF} =0, \\ [10pt]
		\left( {1-\dfrac{r_{0} }{r}} \right)\dfrac{d\tilde{{G}}_{EF} }{dr}+\left( {\dfrac{1-{r_{0} } / r-\kappa /{\sqrt 
					{1+{r_{0} } /r} }}{r}+\dfrac{r_{0} }{4r^{2}}} 
		\right)\tilde{{G}}_{EF} - \\ [10pt]
		- i\dfrac{r_{0} }{r}\left( {E+\dfrac{m}{\sqrt {1+{r_{0} 
					} / r} }} \right)\tilde{{G}}_{EF} +i\left( {\dfrac{1}{1+{r_{0} } / r}\dfrac{r_{0} }{2r^{2}}+\dfrac{r_{0} 
			}{r^{2}}\dfrac{\kappa }{\sqrt {1+{r_{0} } / r} }} \right)\tilde{{F}}_{EF}	+ \\ [10pt]
		+ \left( {E-\dfrac{m}{\sqrt {1+{r_{0} } / r} }} \right)\tilde{{F}}_{EF} =0. \\ 
	\end{array}
\end{equation}
If in the neighborhood of the event horizon of $\left( {r\to r_{0} } 
\right)$, we determine

$\left. {\tilde{{F}}_{EF} } \right|_{r\to r_{0} } =\left( {r-r_{0} } 
\right)^{s}\sum\limits_{k=0}^\infty {\tilde{{f}}_{k} \left( {r-r_{0} } 
	\right)^{k}} ,\,\,\,\left. {\tilde{{G}}_{EF} } \right|_{r\to r_{0} } =\left( 
{r-r_{0} } \right)^{s}\sum\limits_{k=0}^\infty {\tilde{{g}}_{k} \left( 
	{r-r_{0} } \right)^{k}} ,$ then the indicial equation for system (\ref{eq82}) is
\begin{equation}
	\label{eq83}
	s\left( {s+\frac{1}{2}-i2r_{0} E} \right)=0.
\end{equation}
The solutions of (\ref{eq83}) are
\begin{equation}
	\label{eq84}
	s_{1} =-\frac{1}{2}+i2r_{0} E,\,\,\,s_{2} =0.
\end{equation}
The solution of $s_{1} =-1 / 2+i2r_{0} E$ corresponds to asymptotics for 
square-nonintegrable wave functions with the mode of a particle ''fall'' to 
the event horizon. The asymptotics with solution $s_{1} $ coincide with 
asymptotics (\ref{eq78}). These asymptotics agree with the transformations of wave 
function asymptotics $F_{1} ,G_{1} $ (see Eq. (\ref{eq35})) at transition to the EF 
metric and to the Hamiltonian with tetrads in Schwinger gauge.

In this case, the transition to the stationary EF metric does not 
eliminate the mode of a particle ''fall'' to the event horizon. The wave 
functions are square-nonintegrable in the neighborhood of the event horizon.

Since for the (EF) metric, $\eta_{_{EF}} =\left( {1+{r_{0} } /r} \right)^{1 
	/4}$, these conclusions are also valid for the wave functions of the 
self--conjugate Hamiltonian of 
\begin{equation}
	\label{eq85}
	H_{\eta_{_{EF}} } =\eta_{_{EF}} \tilde{{H}}_{_{EF}} \eta_{_{EF}}^{-1} ,
\end{equation}
\begin{equation}
	\label{eq86}
	\psi_{\eta_{_{EF}} } =\eta_{_{EF}} \tilde{{\psi }}_{_{EF}} .
\end{equation}
The smooth square-integrable wave functions of $\left. {\left( 
	{\tilde{{F}}_{EF} } \right)_{2} } \right|_{r\to r_{0} } =\mbox{const}\,\,1$, 
$\left. {\left( {\tilde{{G}}_{EF} } \right)_{2} } \right|_{r\to r_{0} } 
=\mbox{const}\,\,2$ without a mode of a particle ''fall'' to event horizons 
correspond to asymptotics with the solution of $s_{2} =0$ (see Eq. (\ref{eq84})). Some 
of the researches use this solution to prove the possibility of intersection 
of the event horizon by quantum mechanical particles and their ''sink'' to 
singularity at $r\to 0$ (see, for instance, Ref. \cite{bib26}). 
However, the solution of $s_{2} =0$ does not correspond to the history of 
wave function transformations (see (\ref{eq58}), (\ref{eq76}) and (\ref{eq86})) at the transition from the self--conjugate Hamiltonian with the Schwarzschild metric (\ref{eq29}) to the self--conjugate Hamiltonian with the EF metric (\ref{eq85}).

Of course, the wave functions with $s_{2} =0$ are accurate solutions of GR 
equations in the EF space-time. However, these solutions are not 
associated with the Schwarzschild metric. Such an unambiguous link exists 
only for wave functions with the index of $s_{1} =-\frac{1}{2}+i2r_{0} E$ 
(see Eq. (\ref{eq84})).

In accordance with Birkhoff's theorem, the geometry of any spherically 
symmetric vacuum region of spacetime with $r>r_{0} $ is a piece of the 
Schwarzschild geometry. Hence, at coordinate transformations of a static 
Schwarzschild metric to a stationary EF metric, only wave 
functions with an index $s_{1} $ should be used. This leads to conservation 
of the mode of a particle ''fall'' to the event horizon.


\section{The Painlev\`{e}-Gullstrand stationary metric}

For a PG metric, as well as in Sec. 6, we prove the impossibility to 
eliminate the singular mode of a particle ''fall'' by using coordinate 
transformations of the Schwarzschild metric. Because of the logic closeness 
of Secs. 6 and 7, we present a short form of the proofs. 

The coordinate transformation is given by
\begin{equation}
	\label{eq87}
	dT=dt+\sqrt {\frac{r_{0} }{r}} \frac{dr}{f_{S} }.
\end{equation}
The PG metric is written as
\begin{equation}
	\label{eq88}
	ds^{2}=f_{S} dT^{2}-2\sqrt {\frac{r_{0} }{r}} dTdr-dr^{2}-r^{2}\left( 
	{d\theta^{2}+\sin^{2}\theta d\varphi^{2}} \right).
\end{equation}
From a Hamiltonian $\tilde{{H}}_{S} $ in the Schwarzschild space-time (\ref{eq28}), 
the Hamiltonian $H_{PG} $ is obtained in the PG space--time by using 
relationship (\ref{eq19}) and equalities (\ref{eq25}) - (\ref{eq27}). 

As the result, we obtain that
\begin{equation}
	\label{eq89}
	\begin{array}{l}
		\left( {H_{PG} } \right)_{{\underline 0}}^{0} =\dfrac{1}{\sqrt {f_{S} } },\,\,\,\,\left( 
	{H_{PG} } \right)_{{\underline 1}}^{0} =\dfrac{\sqrt {{r_{0} } / r} }{\sqrt 
		{f_{S} } },\,\,\,\,\left( {H_{PG} } \right)_{{\underline 1}}^{1} =\sqrt {f_{S} }, \\ [10pt]
	\left( {H_{PG} } \right)_{{\underline 2}}^{2} 
	=\dfrac{1}{r},\,\,\,\,\,\,\,\left( {H_{PG} } \right)_{{\underline 3}}^{3} 
	=\dfrac{1}{r\sin \theta },
	\end{array}
\end{equation}
\begin{equation}
	\label{eq90}
	\gamma_{PG}^{0} =\frac{1}{\sqrt {f_{S} } }\gamma^{{\underline 0}} +\frac{\sqrt 
		{{r_{0} } / r} }{\sqrt {f_{S} } }\gamma^{{\underline 1}} 
	,\,\,\,\,\gamma_{PG}^{1} =\sqrt {f_{S} } \gamma^{{\underline 1}} ,\,\,\,\gamma 
	_{PG}^{2} =\frac{1}{r}\gamma^{{\underline 2}} ,\,\,\,\,\,\,\,\gamma_{PG}^{3} 
	=\frac{1}{r\sin \theta }\gamma^{{\underline 3}} ,
\end{equation}
\begin{equation}
	\label{eq91}
	\begin{array}{l}
		\left( {\Phi_{PG} } \right)_{0} =\dfrac{r_{0} }{4r^{2}}\gamma^{{\underline 0}} 
		\gamma^{{\underline 1}} ,\,\,\,\,\,\,\,\left( {\Phi_{PG} } \right)_{1} 
		=\dfrac{\sqrt {{r_{0} } / r} }{f_{S} }\dfrac{r_{0} }{4r^{2}}\gamma^{{\underline 0}} 
		\gamma^{{\underline 1}} , \\ [10pt]
		\left( {\Phi_{PG} } \right)_{2} =-\dfrac{1}{2}\sqrt {f_{S} } \gamma 
		^{{\underline 1}} \gamma^{{\underline 2}} ,\,\,\,\,\,\,\,\left( {\Phi_{EF} } \right)_{3} 
		=-\dfrac{1}{2}\cos \theta \,\gamma^{{\underline 2}} \gamma^{{\underline 3}} 
		+\dfrac{1}{2}\sqrt {f_{S} } \sin \theta \gamma^{{\underline 3}} \gamma^{{\underline 1}} . 
		\\ 
	\end{array}
\end{equation}
For the (PG) metric
\begin{equation}
	\label{eq92}
	g^{00}=1.
\end{equation}
By substituting (\ref{eq89}) - (\ref{eq92}) to Eq. (\ref{eq10}), we obtain the Hamiltonian in 
the PG space-time
\begin{equation}
	\label{eq93}
	H_{PG} =\frac{1-\sqrt {{r_{0} } / r} \gamma^{{\underline 0}} \gamma^{{\underline 1}} 
	}{f_{S} }\tilde{{H}}_{S} .
\end{equation}
By analogy with the EF metric (see Eqs. (\ref{eq50}) - (\ref{eq58})), it is possible to 
demonstrate that Hamiltonians $H_{PG} $ and $\tilde{{H}}_{S} $ are connected 
with each other through the unitary transformation:
\begin{equation}
	\label{eq94}
	H_{_{PG}} =e^{iE\varphi_{_{PG}} }\tilde{{H}}_{S} e^{-iE\varphi_{_{PG}} },
\end{equation}
\begin{equation}
	\label{eq95}
	\psi_{_{PG}} \left( {{\rm {\bf r}}} \right)=\tilde{{\psi }}_{_S} \left( {{\rm 
			{\bf r}}} \right)e^{iE\varphi_{_{PG}} },
\end{equation}
where
\begin{equation}
	\label{eq96}
	\varphi_{_{PG}} \left( r \right)=\int {\sqrt {\frac{r_{0} }{r}} 
		\frac{dr}{f_{S} }} .
\end{equation}
Multiplication of asymptotics (\ref{eq39}) and (\ref{eq40}) by the unitary factor of $U_{PG} =e^{iE\varphi_{_{PG}} }$ leads to a form similar to the formulas (\ref{eq60}) and (\ref{eq61}) 
for the EF metric:
\begin{equation}
	\label{eq97}
	\left. {\left( {F_{PG} } \right)_{1} } \right|_{r\to r_{0} } , \,\,\left. {\left( 
		{G_{PG} } \right)_{1} } \right|_{r\to r_{0} } \sim \frac{1}{\left( {r-r_{0} 
		} \right)^{1 / 4}}\exp \left\{ {i2r_{0} E\ln \left( 
		{\frac{r}{r_{0} }-1} \right)} \right\},
\end{equation}
\begin{equation}
	\label{eq98}
	\left. {\left( {F_{PG} } \right)_{2} } \right|_{r\to r_{0} } , \,\,\left. {\left( 
		{G_{PG} } \right)_{2} } \right|_{r\to r_{0} } \sim \frac{1}{\left( {r-r_{0} 
		} \right)^{1 / 4}}.
\end{equation}
The asymptotics (\ref{eq97}) and (\ref{eq98}) can be obtained in a different way. For this, in the equation: 
\begin{equation}
	\label{eq99}
	H_{PG} \psi_{PG} =E\psi_{PG} ,
\end{equation}
it is necessary to separate the variables and obtain the system of equations 
for the radial wave functions $F_{PG} ,G_{PG} $. Then, in the neighborhood 
of the event horizon for this system of equations, we should obtain the 
solutions of the indicial equation. As the result, by analogy with (\ref{eq62}) - 
(\ref{eq64}), we obtain the asymptotics of the radial functions coinciding with 
(\ref{eq97}) and (\ref{eq98}).

The Parker operator in the PG space--time with the set of tetrads (\ref{eq89}) is 
\begin{equation}
	\label{eq100}
	\left( {\rho_{_P} } \right)_{PG} =\gamma^{{\underline 0}}\gamma_{PG}^{0} 
	=\frac{1}{\sqrt {f_{S} } }\left( {1+\sqrt {\frac{r_{0} }{r}} \gamma 
		^{{\underline 0}}\gamma^{{\underline 1}}} \right).
\end{equation}
The operator of transition to the self-conjugate Hamiltonian with a plane 
scalar product is given by
\begin{equation}
	\label{eq101}
	\eta_{_{PG}} =\frac{1}{f_{S}^{1 / 4} }\left( {1+\sqrt {\frac{r_{0} }{r}} \gamma 
		^{{\underline 0}}\gamma^{{\underline 1}}} \right)^{1 / 2}.
\end{equation}
The practical use of a matrix operator $\eta_{_{PG}} $ seems to be difficult.

\subsection{The Lorentz transformation}
\label{sec:mylabel8}
In order to obtain $\eta_{_{PG}} $ in an acceptable matrix-free form, let us 
transfer, as well as in Sec. 6, from the system of tetrad vectors of 
$\left\{ {\left( {H_{PG} } \right)_{{\underline \alpha }}^{\mu } } \right\}$ to tetrad 
vectors of $\left\{ {\left( {\tilde{{H}}_{PG} } \right)_{{\underline \alpha }}^{\mu } } 
\right\}$ in the Schwinger gauge using the Lorentz transformation of (\ref{eq20}) - 
(\ref{eq23}). Non--zero tetrads $\left( {\tilde{{H}}_{PG} } \right)_{{\underline \alpha }}^{\mu 
} $ in the Schwinger gauge are determined by the following expressions:
\begin{equation}
	\label{eq102}
	\begin{array}{l}
	({\tilde H_{PG} })_{{\underline 0}}^{0} =1,\,\,\,\,\,\,( 
	{\tilde{{H}}_{PG} } )_{{\underline 0}}^{1} =-\sqrt {\dfrac{r_{0} }{r}} 
	,\,\,\,\,\,( {\tilde{{H}}_{PG} } )_{{\underline 1}}^{1} =1, \\[10pt]
	( 	{\tilde{{H}}_{PG} } )_{{\underline 2}}^{2} =\dfrac{1}{r},\,\,\,\,\,\,\,( 
	{H_{PG} } )_{{\underline 3}}^{3} =\dfrac{1}{r\sin \theta }.
\end{array}
\end{equation}
For the PG metric, the non--zero values of $\Lambda_{{\underline \alpha }}^{{\underline \beta }} 
\left( r \right)$ in (\ref{eq20}) are
\begin{equation}
	\label{eq103}
	\Lambda_{{\underline 0}}^{{\underline 0}} =\Lambda_{{\underline 1}}^{{\underline 1}} =\frac{1}{\sqrt {f_{S} } 
	},\,\,\,\,\,\Lambda_{{\underline 0}}^{{\underline 1}} =\Lambda_{{\underline 1}}^{{\underline 0}} =-\frac{\sqrt {{r_{0} 
			} /r} }{\sqrt {f_{S} } }.
\end{equation}
Similarly to Eqs. (\ref{eq69}) - (\ref{eq74}), we obtain the following form of a matrix $L,L^{-1}$ for the PG metric:
\begin{equation}
	\label{eq104}
	L\left( r \right)=\frac{1+\sqrt {{r_{0} }/ r} +\sqrt {f_{S} } +\left( {1+\sqrt 
			{{r_{0} } / r} -\sqrt {f_{S} } } \right)\gamma^{{\underline 0}}\gamma 
		^{{\underline 1}} }{2\sqrt {\sqrt {f_{S} } \left( {1+\sqrt {{r_{0} } / r} } \right)} 
	},\,
\end{equation}
\begin{equation}
	\label{eq105}
	L^{-1}\left( r \right)=\frac{1+\sqrt {{r_{0} } / r} +\sqrt {f_{S} } +\left( 
		{1+\sqrt {{r_{0} } / r} -\sqrt {f_{S} } } \right)\gamma^{{\underline 1}} 
		\gamma^{{\underline 0}}}{2\sqrt {\sqrt {f_{S} } \left( {1+\sqrt {{r_{0} } 
					/r} } \right)} }.
\end{equation}
After Lorentz transformation (\ref{eq104}) and (\ref{eq105}), Hamiltonian (\ref{eq94}) is transformed to a self--conjugate Hamiltonian with tetrads in the Schwinger gauge \cite{bib14}, \cite{bib26}.

For PG metrics, $\eta_{_{PG}} =1$ and $\tilde{{H}}_{PG} =H_{\eta_{_{PG}} } $, 
where 
\begin{equation}
	\label{eq106}
	\begin{array}{l}
		H_{\eta_{_{PG}} } =L\left( r \right)H_{PG} L^{-1}\left( r \right)= \\ [10pt]
		=m\gamma^{{\underline 0}}-i\gamma^{{\underline 0}}\left\{ {\gamma^{{\underline 1}}\left( {\dfrac{\partial 
				}{\partial r}+\dfrac{1}{r}} \right)+\gamma^{{\underline 2}}\dfrac{1}{r}\left( 
			{\dfrac{\partial }{\partial \theta }+\dfrac{1}{2}\mbox{ctg}\theta } 
			\right)+\gamma^{{\underline 3}}\dfrac{1}{r\,\mbox{sin}\theta }} \right\}+ \\ [10pt]
		+ i\sqrt {\dfrac{r_{0} }{r}} \left( {\dfrac{\partial }{\partial 
				r}+\dfrac{3}{4}\dfrac{1}{r}} \right), \\ [10pt]
		\psi_{\eta_{_{PG}} } =L\left( r \right)\psi_{PG} . \\ 
	\end{array}
\end{equation}
According to (\ref{eq104}), the asymptotics $L\left( r \right)$ at $r\to r_{0} $ 
equals
\begin{equation}
	\label{eq107}
	\left. L \right|_{r\to r_{0} } =\frac{r_{0}^{1 / 4} \left( {1+\gamma^{{\underline 0}}\gamma 
			^{{\underline 1}} } \right)}{2^{1 / 2}\left( {r-r_{0} } \right)^{1 / 4}}.
\end{equation}
In compliance with (\ref{eq106}) and (\ref{eq107}), after the similarity transformation 
$L\left( r \right)$, the asymptotics of radial functions (\ref{eq97}), (\ref{eq98}) become equal to
\begin{equation}
	\label{eq108}
	\left. {\left( {\tilde{{F}}_{pG} } \right)_{1} } \right|_{r\to r_{0} } 
	, \,\,\left. {\left( {\tilde{{G}}_{PG} } \right)_{1} } \right|_{r\to r_{0} } \sim 
	\frac{1}{\left( {r-r_{0} } \right)^{1 / 2}}\exp \left\{ {i2r_{0} E\ln \left( 
		{\frac{r}{r_{0} }-1} \right)} \right\},
\end{equation}
\begin{equation}
	\label{eq109}
	\left. {\left( {\tilde{{F}}_{PG} } \right)_{2} } \right|_{r\to r_{0} } 
	, \,\,\left. {\left( {\tilde{{G}}_{PG} } \right)_{2} } \right|_{r\to r_{0} } \sim 
	\frac{1}{\left( {r-r_{0} } \right)^{1 / 2}}.
\end{equation}
Then, we will obtain asymptotics of radial functions in a different way.

While presenting a wave function $\psi_{\eta_{_{PG}} } \left( {{\rm {\bf r}},T} \right)$ in the form of (\ref{eq30}) in the equation of 
\begin{equation}
	\label{eq110}
	H_{\eta_{_{PG}} } \psi_{\eta_{_{PG}} } =E\psi_{\eta_{_{PG}} } 
\end{equation}
we can separate the variables. While separating the variables, as well as in 
Secs. 5 and 6, let us perform an equivalent substitution of matrixes $\gamma 
^{{\underline i}}$ (see Eq. (\ref{eq31})). 

The system of equations for radial wave functions $\tilde{{F}}_{PG} \left( r 
\right),\tilde{{G}}_{PG} \left( r \right)$ is given by
\begin{equation}
	\label{eq111}
	\begin{array}{l}
		f_{S} \dfrac{d\tilde{{F}}_{PG} }{dr}+\left( {\dfrac{1+\kappa 
			}{r}-\dfrac{3}{4}\dfrac{r_{0} }{r^{2}}-i\left( {E-m} \right)\sqrt {\dfrac{r_{0} 
				}{r}} } \right)\tilde{{F}}_{PG} - \\ [10pt]
			- \left( {E+m+i\sqrt {\dfrac{r_{0} }{r}} 
			\dfrac{1 /4-\kappa }{r}} \right)\tilde{{G}}_{PG} =0, \\ [10pt]
		f_{S} \dfrac{d\tilde{{G}}_{PG} }{dr}+\left( {\dfrac{1-\kappa 
			}{r}-\dfrac{3}{4}\dfrac{r_{0} }{r^{2}}- i\left( {E+m} \right)\sqrt {\dfrac{r_{0} 
				}{r}} } \right)\tilde{{G}}_{PG} + \\ [10pt] 
			+ \left( {E-m+i\sqrt {\dfrac{r_{0} }{r}} 
			\dfrac{1 /4+\kappa }{r}} \right)\tilde{{F}}_{PG} =0 \\ 
	\end{array}
\end{equation}
The indicial equation for system (\ref{eq111}) coincides with equation (\ref{eq83}) for the (EF) metric.

The solution of $s_{1} =-1 / 2+i2r_{0} E$ corresponds to square-nonintegrable 
wave functions with the mode of a particle ''fall'' to the event horizon. 
This solution corresponds to the wave function transformations at transition 
to the PG metric.

The solution of $s_{2} =0$ does not correspond to the transformation history 
of wave functions at transition from the Schwarzschild metric to the PG 
metric. As well as for the EF metric, the solution of $s_{2} =0$ for the 
PG metric, being an accurate solution of the GR equations, is not 
connected with the Schwarzschild solution.

The connection with the original Schwarzschild metric is implemented only by 
the solution with the index of $s_{1} =-1 /2+i2r_{0} E$.

It follows therefrom that the transition to the (PG) stationary metric does 
not eliminate the mode of a particle ''fall'' to the event horizon. The wave 
functions are square-nonintegrable in the neighborhood of the event horizon.


\section{Static metrics and the Schwarzschild metric with ''tortoise'' coordinate}

At spatial coordinate transformations of the Schwarzschild metric, the 
radius of the event horizon is changed. For instance, for the metric in 
isotropic coordinates \cite{bib27, bib28}, the gravitational radius 
is $\left( {R_{is} } \right)_{0} ={r_{0} } / 4$, for the metric in 
spherical harmonic coordinates, $\left( {R_{g} } \right)_{0} ={r_{0} } 
/ 2$. In this case, the mode of a particle ''fall'' on the event horizons is 
preserved (see, in Ref. \cite{bib14} solutions (\ref{eq35}) and (\ref{eq41}), 
Hamiltonian (\ref{eq61}) and the system of equations for radial functions (\ref{eq62})).

The similar situation is preserved when using the transformation of Ref. \cite{bib10}
\begin{equation}
	\label{eq112}
	dr\ast =dr\left( {1-\frac{r_{0} }{r}} \right)^{-1},
\end{equation}
where
\begin{equation}
	\label{eq113}
	r\ast =r+r_{0} \mbox{ln}\left( {\frac{r_{0} }{r}-1} \right)+\mbox{const.}
\end{equation}
A value $r$ is the function of $r\ast $.

The coordinate $r\ast \to -\infty $ at $r\left( {r\ast } \right)\to r_{0} $. 
In the system of equations (\ref{eq32}), for the Schwarzschild metric at 
transformation (\ref{eq112}), the first summands become equal to ${dF} 
/{dr\ast },\,\,{dG} /{dr\ast }$. In the rest of the summands, the function of $r\left( {r\ast } \right)$ is used. 
Asymptotically, at $r\ast \to -\infty $ and $r\left( {r\ast } \right)\to 
r_{0} $, the solutions of system (\ref{eq32}) are given by
\begin{equation}
	\label{eq114}
	\begin{array}{l}
		\left. F \right|_{r\ast \to -\infty } =L\ast \left( {r\left( {r\ast } 
			\right)-r_{0} } \right)^{{-1} /2}\sin \left( {Er\ast +\delta \ast } \right), \\ [10pt]
		\left. G \right|_{r\ast \to -\infty } =L\ast \left( {r\left( {r\ast } 
			\right)-r_{0} } \right)^{{-1} / 2}\cos \left( {Er\ast +\delta \ast } \right), \\ 
	\end{array}
\end{equation}
where $L\ast ,\delta \ast $ are integration constants. 

The wave functions in (\ref{eq114}) preserve the mode of a particle ''fall'' and are 
square-nonintegrable (see also Eq. (\ref{eq37})).

\section{Nonstationary metrics}

After transformations from the Schwarzschild metric to the space-time of 
Lema\'{\i}tre--Finkelstein and Kruskal-Szekeres nonstationary metrics, Dirac Hamiltonians 
explicitly depend on time coordinate \cite{bib14}. In these 
cases, examination of stationary states of particles is impossible \cite{bib4}.

\section{Conclusions}
The coordinate transformations of the Schwarzschild metric do not eliminate 
the singular mode of a particle ''fall'' to the event horizon of a black 
hole. This mode is unacceptable for quantum mechanics of stationary states.

This conclusion has been rigorously proved for EF and 
PG stationary metrics and is also valid for any 
transformed static metric, including Schwarzschild metric with a 
''tortoise'' coordinate $r\ast $.


\section*{Acknowledgments}

The author thanks M.~A.~Vronskiy and V~.E.~Shemarulin for the useful remarks and 
discussions. The author also thanks L.~P.~Babich and A.~L.~Novoselova for the 
essential technical assistance in preparation of the paper.



\end{document}